\documentclass[twoside,twocolumn]{article}

\usepackage{blindtext} % Package to generate dummy text throughout this template 

\usepackage[sc]{mathpazo} % Use the Palatino font
\usepackage[scale=0.93]{tgpagella} % Adjusting font size
\usepackage[T1]{fontenc} % Use 8-bit encoding that has 256 glyphs
\usepackage{microtype} % Slightly tweak font spacing for aesthetics

\usepackage[english]{babel} % Language hyphenation and typographical rules

\usepackage[hmarginratio=1:1,top=32mm,columnsep=20pt]{geometry} % Document margins
\usepackage[hang, small,labelfont=bf,up,textfont=it,up]{caption} % Custom captions under/above floats 
\usepackage{float}
\usepackage{booktabs} % Horizontal rules in tables

\usepackage{lettrine} % The lettrine is the first enlarged letter at the beginning of the text

\usepackage{enumitem} % Customized lists
\setlist[itemize]{noitemsep} % Make itemize lists more compact

\usepackage{abstract} % Allows abstract customization
\usepackage{titlesec} % Allows customization of titles
\renewcommand\thesection{\Roman{section}} % Roman numerals for the sections
\renewcommand\thesubsection{\roman{subsection}} % roman numerals for subsections
\titleformat{\section}[block]{\large\scshape\centering}{\thesection.}{1em}{} % Change the look of the 
\titleformat{\subsection}[block]{}{\thesubsection.}{1em}{} % Change the look of the section titles

\usepackage{fancyhdr} % Headers and footers
\usepackage{titling} % Customizing the title section

\usepackage[dvipsnames]{xcolor}
\usepackage{hyperref}

\colorlet{mylinkcolor}{Black}
\colorlet{mycitecolor}{Black}
\colorlet{myurlcolor}{Blue}
\usepackage{graphicx}
\usepackage{cite}

\hypersetup{
  linkcolor  = mylinkcolor,
  citecolor  = black,
  urlcolor   = myurlcolor,
  colorlinks = true,
}

\usepackage{graphicx}

\newcommand{\ORCIDDisplay}[1]{\href{http://orcid.org/\#1}{\includegraphics[scale=.10]{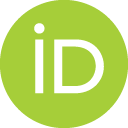}}}
\newcommand{\firstAuthor}{Gopal P. Sarma}
\newcommand{\firstAuthorEmail}{gopal.sarma@emory.edu}
\newcommand{\firstAuthorAffiliation}{School of Medicine, Emory University, Atlanta, GA USA}
\newcommand{\firstAuthorORCID}{0000-0002-9413-6202}

\pretitle{\begin{center}\huge\bfseries} % Article title formatting
\posttitle{\end{center}} % Article title closing formatting
\title{Scientific Auditing Firms} % Article title
\author{
\textsc{
\firstAuthor\ORCIDDisplay{\firstAuthorORCID}\textsuperscript{1}\thanks{Email: \firstAuthorEmail}\hspace{2pt} 
} \vspace{8pt} \\ % Your name
\normalsize 1. \emph{\firstAuthorAffiliation}\\ % Your institution
}
\date{} % Leave empty to omit a date
\renewcommand{\maketitlehookd}{%
\begin{abstract}
\noindent The ``crisis of reproducibility'' has been a significant source of controversy, heated debate, and calls for reform to institutional science in recent years.  As a long-term solution to address both the present crisis and future obstacles, I propose the creation of a new form of research organization whose purpose would be to conduct random audits of the scientific literature.   I suggest that data analytics of a digitized scientific corpus may play a critical role in allowing broadly educated scientists to identify linchpin results to investigate in further detail across all disciplines.  I argue that a simple ``mock'' trial run of a simplified auditing firm consisting of several researchers over a short time period would provide valuable insight into the feasibility of this proposal.
\end{abstract}
}

\begin{document}

% Print the title
\maketitle

%----------------------------------------------------------------------------------------
%	ARTICLE CONTENTS
%----------------------------------------------------------------------------------------

\section{Introduction}
\setlength{\parindent}{0cm}
[\emph{A longer version of this article is available via \href{http://www.pubpub.org/pub/scientific-auditing-firms}{PubPub}}] \\

\emph{\textbf{Can non-specialists with advanced scientific training identify key results worth replicating in a field that they have little to no experience with?}}  If so, there would be profound consequences for the future of science.  In particular, it would allow for the creation of a new form of research organization, which I term \emph{scientific auditing firms}.  Their primary responsibility would be to conduct random, systematically identified audits of the scientific literature.  In addition to creating a disincentive for those who might otherwise engage in fraudulent practices, the existence of full-time, independent auditing firms would give academia and industry a greater sense of security in the reliability of the scientific corpus.  \\

In such an organization, there would be a concentration of outstanding scientists exposed to the breadth of research produced by the entire scientific establishment.  This fact would have significant secondary implications.  For instance, auditing firms might also come to play the role of global monitors of scientific progress, issuing regular technical reports on contemporary developments, collaborating with filmmakers to develop documentaries of particular importance to the public, or offering technical consulting services for academia and industry. \\

Why is it essential that these results be identifiable by non-experts?  The explosive growth of the scientific enterprise following the Second World War has paralleled a trend towards hyper-specialization.  Consequently, a thorough understanding of a given research result almost always requires extensive training in a specific field.  It would not be possible, therefore, for an organization to employ specialists from every field \cite{Bode03061949, sarma1, sarma2, sarma3} .   \\

We have, however, a powerful set of tools that have only recently come into existence, namely, a digitized scientific corpus and the techniques of modern data science \cite{markowitz2015linguistic, ding2011applying, ding2011topic, ding2013distribution, zhu2013bibliometric, zhu2015measuring, song2014productivity, valverde2007topology, gress2010properties, solee2013evolutionary}.  For analyzing the scientific corpus, relevant data science techniques include citation network analysis, natural language processing, and many other statistical methods developed for the processing of large data sets.  Using these tools, it may be possible for an experienced scientist with strong quantitative skills to identify those experiments or results that merit further investigation and which lie outside of the scope of their scientific training \cite{Sarma2017Scientific}.  The identification of such results would constitute the first step of conducting a ``scientific audit.''  \\

Subsequent steps---which would require the participation of specialists from the field in question---might range from full-scale replication of an experiment, to the writing of a review article or set of tutorials on novel statistical techniques, to coordinating the investigation of a result with alternate methods via a network of collaborating laboratories.  \\

The ultimate consequences of random scientific audits would be more than intellectual.  Although it is difficult to quantify, the reproducibility crisis has come with a steep cost to science, industry, and society as a whole.  The combination of poor or outright fraudulent research has resulted in significantly wasted financial resources, much of which has come from the tax-paying public.  In addition to the cost of delayed scientific and technological development, there is now the additional cost of investigating and characterizing the severity of the problem itself.  The recent analyses that have revealed large numbers of problematic studies were in a limited range of subjects and we can hardly claim to know what this distribution looks like for the entirety of science \cite{Sarma2017Scientific, Campbell2015a, Ioannidis2005, GoodmanAndGreeland, Horton2015, Prinz2011, Alberts22042014, gunn2014reproducibility, adam2002journals, check2005korean, bouri2014meta}.

\section{Evaluating the Feasibility of Independent Auditing Firms}
As described above, the fundamental notion of a scientific auditing firm is quite simply stated.  It would be a completely neutral organization, with no research objectives of its own, whose primary purpose would be to conduct random, systematically identified audits of the scientific literature.  Nevertheless, the practicalities of how such an organization would operate, its relationship to the university system, and the network of relationships that would be required to conduct an audit are likely to be quite complex and involve many subtleties which we cannot currently anticipate.  \\

Therefore, in order to evaluate the feasibility of establishing full-fledged, independent scientific auditing firms, I propose that we take an empirical stance and conduct a simple experiment to answer the question that motivated this article: \emph{\textbf{Can non-specialists with advanced scientific training identify key results worth replicating in a field that they have little to no experience with?}}  The experiment would consist of funding 1-2 researchers with broad scientific training and data science experience to conduct a ``mock'' trial run of an auditing firm.  The goal would be to understand the challenges for non-experts to identify critical results to investigate in fields outside of their direct scientific training.  For this initial experiment, we would not need to proceed with the auditing process itself.  Simply understanding the challenges of the identification phase would be valuable.  \\

There are many lessons we would learn from conducting such an experiment, ranging from the skills and experience that would be required of scientific auditors, to limitations on current data science toolkits for analyzing the scientific corpus, to the value of old-fashioned ``investigative journalism'' in the auditing process.  We would also be forced to confront issues related to open-access of the scientific literature and whether partial availability of the research corpus in a given discipline is sufficient to reliably identify linchpin results.  \\

The reproducibility crisis is a deeply troubling development that should motivate us to think critically and creatively about the future health of institutional science.  In addition to the many reforms being proposed today, scientific auditing firms merit serious consideration as a long-term solution to ensure the reliability of published results.  While there is much to be gained by discussing the practicalities and nuances of this proposal, there is a fundamental question which I have stated above that we can evaluate empirically.  It would cost very little to conduct a ``mock'' trial run of a simplified auditing firm and the outcome of this experiment would inform whether further consideration of this idea is merited.  

\subsection*{Acknowledgements}
I would like to acknowledge Seshu Sarma and Caroline Schwenz for feedback on the manuscript. 

\section*{ORCID}
\makebox[2.5cm]{\firstAuthor} \raisebox{-.26\height}{\includegraphics[scale=.10]{orcid128}} \href{http://orcid.org/\firstAuthorORCID}{\firstAuthorORCID}\\

\bibliographystyle{ieeetr}
\bibliography{scientific_auditing_firms_short}

\end{document}